\documentclass[twocolumn,conference]{IEEEtran}
\usepackage[pdftex]{graphicx}
\usepackage[left=0.625in, right=0.625in, top=0.75in, bottom=1in]{geometry}

\usepackage{bm}
\usepackage{dsfont}
\usepackage{mathrsfs}
\usepackage{authblk}
\usepackage{amsmath}
\usepackage{graphicx}
\usepackage{amssymb}
\usepackage{epstopdf}
\usepackage{subfigure}
\usepackage{textcomp}
\usepackage{color}
\usepackage{upgreek}
\usepackage{cite}
\usepackage{cleveref}
\usepackage[utf8]{inputenc}
\usepackage[english]{babel}
\usepackage{algorithm,algpseudocode}

\DeclareMathOperator*{\argmax}{arg\,max}
\DeclareMathOperator*{\argmin}{arg\,min}

\begin{document}
\title{A Low-Complexity Cache-Aided Multi-Antenna Content Delivery Scheme}
\author{Junlin Zhao}
\author{Mohammad Mohammadi Amiri}
\author{Deniz G\"und\"uz}
\affil{Department of Electrical and Electronic Engineering, Imperial College London, London, SW7 2BT, UK
\authorcr Emails: \{j.zhao15, m.mohammadi-amiri15, d.gunduz\}@imperial.ac.uk}
\maketitle

\begin{abstract}

We study downlink beamforming in a single-cell network with a multi-antenna base station (BS) serving cache-enabled users.
For a given common rate of the files in the system, we first formulate the minimum transmit power with beamforming at the BS as a non-convex optimization problem. This corresponds to a multiple multicast problem, to which a stationary solution can be efficiently obtained through successive convex approximation (SCA). It is observed that the complexity of the problem grows exponentially with the number of subfiles delivered to each user in each time slot, which itself grows exponentially with the number of users in the system. Therefore, we introduce a low-complexity alternative through time-sharing that limits the number of subfiles that can be received by a user in each time slot.
It is shown through numerical simulations that, the reduced-complexity beamforming scheme has minimal performance gap compared to transmitting all the subfiles jointly, and outperforms the state-of-the-art low-complexity scheme at all SNR and rate values with sufficient spatial degrees of freedom, and in the high SNR/high rate regime when the number of spatial degrees of freedom is limited.\makeatletter{\renewcommand*{\@makefnmark}{}\footnotetext{This work was partially supported by the European Research Council (ERC) through project BEACON (No. 725731), and by the European Union’s Horizon 2020 Research and Innovation Programme through project SCAVENGE (No. 675891).}\makeatother}
\end{abstract}
%\IEEEpeersolutionreviewmaketitle
\IEEEpeerreviewmaketitle

\section{Introduction}

Research efforts in efficient video delivery have intensified recently.
A promising approach to enhance the performance of content delivery is to exploit cache memories across the network. Proactive content caching at user devices during off-peak periods can improve the spectral efficiency, and reduce the bandwidth requirements and the transmission delay, particularly for video-on-demand services \cite{AlmerothCacing}.

The coded cache placement and delivery scheme proposed in \cite{6763007} creates and exploits multicasting opportunities to users with arbitrary demands; thereby achieving a global caching gain. While \cite{6763007} considered a noiseless broadcast channel, application of coded delivery in real cellular systems requires designing a multicasting strategy over a noisy broadcast channel jointly with the caching and delivery scheme \cite{DBLP:journals/corr/abs-1711-05969,MohammadDenizJSACPower}.
As an efficient physical layer transmission technique,
multicast beamforming, where the multi-antenna base station (BS) multicasts distinct data streams to multiple user groups, has been extensively studied in cellular networks \cite{Luo,Xiang}. Inspired by the multicasting feature of coded content delivery schemes \cite{6763007}, and the high efficiency of multi-antenna beamforming, there have been recent research efforts to incorporate multi-antenna beamforming techniques into cache-aided coded content delivery. In \cite{DBLP:journals/corr/abs-1711-05969}, coded delivery is employed along with zero-forcing to simultaneously exploit spatial multiplexing and caching gains. By treating the transmission of coded subfiles as a coordinated beamforming problem, improved spectral efficiency is achieved in \cite{DBLP:journals/corr/abs-1711-03364} by optimizing the beamforming vectors. A reduced-complexity scheme is also presented in \cite{DBLP:journals/corr/abs-1711-03364} by limiting the number of users served at any time.

\begin{figure}[!t]
\centering
\includegraphics[width=2.65in]{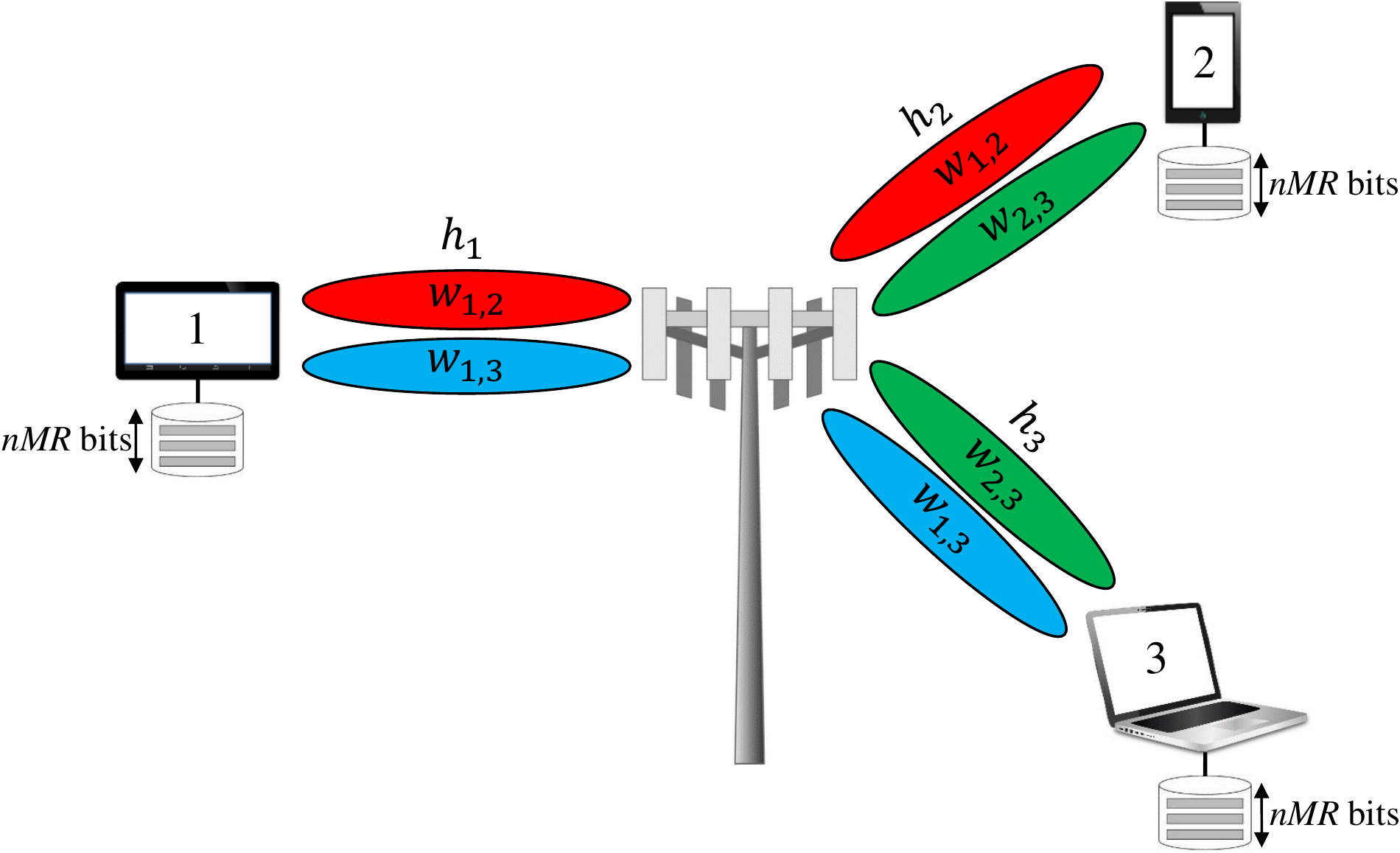}
\caption{Illustration of the cache-aided content delivery system with beamforming for $K=3$ users.}
\label{Sys_Mod}
\end{figure}

In this paper, we consider the problem studied in \cite{DBLP:journals/corr/abs-1711-03364}.
Firstly, a general framework for cache-aided downlink beamforming is formulated, focusing on the minimum required transmit power under user quality-of-service (QoS) requirements. The resultant nonconvex optimization problem is tackled by successive convex approximation (SCA), which is guaranteed to converge to a stationary solution of the original nonconvex problem.
As noted in \cite{DBLP:journals/corr/abs-1711-03364}, the transceiver design and computational complexities increase drastically with the number of coded messages each user decodes at any time slot. We propose a novel content delivery scheme which flexibly adjusts the number of coded messages each user decodes in each time slot.
Unlike the scheme in \cite{DBLP:journals/corr/abs-1711-03364}, our proposed scheme does not limit the number of users served in each time slot, but directly limits the number of messages each user decodes. Our numerical results show that the proposed scheme can provide significant gains in terms of transmit power, particularly in the high rate/high signal-to-noise ratio (SNR) regime.

\section{System Model}\label{SystemModelSec}

%\subsection{System Model}

We consider downlink transmission within a single cell, where a BS equipped with $N_t$ antennas serves $K$ single-antenna users, as illustrated in Fig.~\ref{Sys_Mod}. A library of $N$ files $ \boldsymbol{V} \triangleq (V_1,\cdots,V_N)$, each distributed uniformly over set $\{ 1, \dots, 2^{nR} \} \triangleq \left[ 2^{nR} \right]$ is available at the BS, where $R$ and $n$ represent the rate of each file and the blocklength, respectively. Each user is equipped with a cache that can store up to $M$ files, and the \textit{caching factor} is defined as $t \triangleq MK/N$.

Contents are placed at users' caches during off-peak periods, known as the \textit{placement phase}, without any prior information on user requests. The caching function for user $k$ is denoted by $\phi_k^{(n)} : \left[ 2^{nR} \right]^N \to \left[ 2^{nMR} \right]$, which maps the library to the cache contents $Z_k$ at user $k$, i.e., $Z_k = \phi_k (\boldsymbol{V})$, $k \in [K]$. After the users reveal their demands $\boldsymbol{d} \triangleq (d_1, \dots, d_K)$, signal $\boldsymbol{x} \in \mathbb{C}^{N_t \times n}$ is transmitted by the BS, where we denote $\bm{x} = [\bm{x}_1 \cdots \bm{x}_n]$, $\bm{x}_i \in \mathbb{C}^{N_t \times 1}$. An average power constraint $P$ is imposed on each channel input $\bm{x}$. User $k$ receives
\begin{align}\label{UserkSysModel}
\boldsymbol{y}_k = \boldsymbol{h}_k^H \boldsymbol{x} + \boldsymbol{n}_k,    
\end{align}
where $\bm{h}_k \in \mathbb{C}^{N_t \times 1}$ is the channel vector from the BS to the $k$-th user, and $\bm{n}_k \in \mathbb{C}^{1 \times n}$ is the additive white Gaussian noise at user $k$ with each entry independent and identically distributed according to $\mathcal{CN}(0,\sigma_k^2)$, $k \in [K]$. We assume the channel state information (CSI) is perfectly known to the BS and the receivers. Hence, the encoding function is $\psi^{(n)}: {\left[ 2^{nR} \right]^N} \times {\left[ N \right]^K} \times \mathbb{C}^{N_t \times K} \to \mathbb{C}^{N_t \times n}$. After receiving $\boldsymbol{y}_k$, user $k$ reconstructs $\hat{V}_{d_k}$ using its cache content $Z_k$, channel vector $\bm{h}_k$, and demand vector $\bm{d}$ through function $\mu_k^{(n)}: \mathbb{C}^n \times \left[ 2^{nMR} \right] \times \mathbb{C}^{N_t \times 1} \times \left[ N \right]^K \to \left[ 2^{nR} \right]$, i.e., $\hat{V}_{k} = \mu^{(n)}_k \left( \boldsymbol{y}_k, Z_k, \bm{h}_k, \bm{d} \right)$, $k \in [K]$. The probability of error is defined as ${\rm{P_e}} \triangleq \max_{\bm{d}} \max_{k \in [K]} {\rm{Pr}} \{ V_{d_k} \ne \hat{V}_k \}$. 
An $(R,M,P)$ tuple is \textit{achievable} if there exist a sequence of caching functions ${\phi_1^{(n)}, \dots, \phi_K^{(n)}}$, encoding function $\psi^{(n)}$, and decoding functions ${\mu_1^{(n)}, \dots, \mu_K^{(n)}}$, such that ${\rm{P_e}} \to 0$ as $n \to \infty$. For file rate $R$ and cache size $M$, our main goal is to characterize 
\begin{align}\label{AveragePowerMemoryTradeOff}
P^*\left( {R,M} \right) \buildrel \Delta \over = &\inf \left\{ { P:\left( {R,M, P} \right) \mbox{is achievable}} \right\}.
\end{align}

\section{An Achievable Delivery Scheme}
In this section, we present a multi-antenna transmission scheme with coded caching, following the placement and coded content generation proposed in \cite{6763007}.

\subsection{Placement and Delivery Schemes}
We follow the cache placement scheme of \cite{6763007}: For a caching factor $t \in \{ 1, \dots, K-1 \}$, we represent $t$-element subsets of $[K]$ by $\mathcal{G}^t_1, \dots, \mathcal{G}^t_{\binom{K}{t}}$. File $V_i$, $i \in [N]$, is divided equally into $\binom{K}{t}$ disjoint subfiles $V_{i,\mathcal{G}^t_1},\dots, V_{i,\mathcal{G}^t_{\binom{K}{{t}}}}$, each of $n\frac{R}{\binom{K}{t}}$ bits. User $k$, $k \in [K]$, caches subfile $V_{i,\mathcal{G}^t_j}$, if $k \in \mathcal{G}^t_j$, $\forall j \in [ \binom{K}{t} ]$. The cache content of user $k$ is then given by $\bigcup\nolimits_{i \in [N]} {\bigcup\nolimits_{j \in [\binom{K}{t}]: k \in \mathcal{G}^t_j} {V_{i,\mathcal{G}^t_j}} }$.

The \textit{delivery phase} is performed once the demands are revealed.
For any demand combination $\bm{d}$, we aim to deliver the coded message  
\begin{equation}\label{DefCodedDeliveredContentCentralized}
s_{\mathcal{G}_j^{{t}+1}} \buildrel \Delta \over = {{\bigoplus}_{k \in \mathcal{G}_j^{{t}+1}} {V_{{d_k},\mathcal{G}_j^{{t}+1}\backslash \{ k\} }}} 
\end{equation}
to all the users in set ${\mathcal{G}_j^{{t}+1}}$, for $j \in [ \binom{K}{t+1} ]$. Observe that, after receiving $s_{\mathcal{G}_j^{{t}+1}}$, each user $k \in \mathcal{G}_j^{{t}+1}$ can recover subfile $V_{{d_k},\mathcal{G}_j^{{t}+1}\backslash \{ k\} }$ having access to $V_{{d_l},\mathcal{G}_j^{{t}+1}\backslash \{ l\} }$, $\forall l \in \mathcal{G}_j^{{t}+1} \backslash \{k\}$.

We define $\mathcal{S} \triangleq \{\mathcal{G}^{t+1}_1, \dots, \mathcal{G}^{t+1}_{\binom{K}{t+1}}\}$ as the set of all the messages, with each message $\mathcal{T}\in\mathcal{S}$ represented by the set of users it is targeting, and $\mathcal{S}_k \subset \mathcal{S}$ is the subset of messages targeting user $k$. We have $\left| \mathcal{S} \right| = \binom{K}{t+1}$ and $\left| \mathcal{S} _k\right| = \binom{K-1}{t}$.

The following example will be used throughout the paper to explain the proposed scheme:
\\
\textbf{\textit{Example}}: Let $N=5,~K=5,~M=1$.\label{SubSecExample2}~We have $t=\frac{MK}{N}=1$. Each file is split into $\binom{K}{t}=5$ disjoint subfiles of the same size, where we represent file $i$, $i \in [N]$, as
\begin{align}
V_i = \left\{ V_{i, \{ 1 \}}, V_{i, \{ 2 \}}, V_{i, \{ 3 \}} V_{i, \{ 4 \}},V_{i, \{ 5 \}}\right\}.
\end{align}
The cache content of user $k$ is $Z_k = \cup_{n \in [N]} V_{n, \{k\}}$, $k \in [K]$,
which satisfies the cache capacity constraint. For a demand combination $\bm{d}$, all user demands can be fulfilled by delivering the following $\binom{K}{t+1}=10$ subfiles:
% {
% \small{
\begin{align}
&s_{\{1,2\}} = V_{d_1,\{ 2 \}} \oplus V_{d_2,\{ 1 \}},~s_{\{1,3\}} = V_{d_1,\{ 3 \}} \oplus V_{d_3,\{ 1 \}},\nonumber\\
&s_{\{1,4\}} = V_{d_1,\{ 4 \}} \oplus V_{d_4,\{ 1 \}},~s_{\{1,5\}} = V_{d_1,\{ 5 \}} \oplus V_{d_5,\{ 1 \}},\nonumber\\
&s_{\{2,3\}} = V_{d_2,\{ 3 \}} \oplus V_{d_3,\{ 2 \}},~s_{\{2,4\}} = V_{d_2,\{ 4 \}} \oplus V_{d_4,\{ 2 \}},\nonumber\\
&s_{\{2,5\}} = V_{d_2,\{ 5 \}} \oplus V_{d_5,\{ 2 \}},~s_{\{3,4\}} = V_{d_3,\{ 4 \}} \oplus V_{d_4,\{ 3 \}},\nonumber\\
&s_{\{3,5\}} = V_{d_3,\{ 5 \}} \oplus V_{d_5,\{ 3 \}},~s_{\{4,5\}} = V_{d_4,\{ 5 \}} \oplus V_{d_5,\{ 4 \}}.\nonumber
\end{align}
% }
\subsection{Multi-antenna Transmission Scheme}

The delivery of the coded messages in set $\mathcal{S}$ to their respective sets of receivers is a multi-antenna multi-message multicasting problem. Before introducing our low-complexity scheme in the next section, we present here a general transmission strategy based on message-splitting and time-division transmission. The messages in $\mathcal{S}$ are transmitted over $B$ orthogonal time slots, the $i$-th of which is of
blocklength $n_i$, $i \in [B]$, where $\sum\nolimits_{i=1}^{N} n_i = n$.
The transmitted signal $\bm{x}(i) \triangleq [\bm{x}_{\sum_{j=1}^{i-1}n_j+1} \cdots \bm{x}_{\sum_{j=1}^{i}n_j}  ]$ at time slot $i \in [B]$ is given as
\begin{align}
    \bm{x}(i) = \sum\nolimits_{\mathcal{T} \in \mathcal{S}} \bm{w}_{\mathcal{T}}(i) \bm{s}_\mathcal{T} (i),
\end{align}
where $\bm{s}_\mathcal{T} (i)  \in \mathbb{C}^{1 \times n_i}$ is the unit power complex Gaussian signal of block length $n_i$, modulated from the corresponding message $s_\mathcal{T}$ in (\ref{DefCodedDeliveredContentCentralized}), intended for the users in set $\mathcal{T}$, transmitted in time slot $i$, encoded by the beamforming vector $\bm{w}_\mathcal{T}(i)\in \mathbb{C}^{N_t \times 1}$.

The received signal at user $k$ in time slot $i$ is
\begin{align}\label{receivedSigUserk}
&\bm{y}_k(i) =\nonumber\\
& \; \underbrace{\bm{h}_k^H \sum\limits_{\mathcal{T} \in \mathcal{S}_k} \bm{w}_{\mathcal{T}}(i) \bm{s}_\mathcal{T} (i)}_{\text{desired messages}} + \underbrace{\bm{h}_k^H \sum\limits_{\mathcal{I} \in \mathcal{S}_k^C} \bm{w}_{\mathcal{I}}(i) \bm{s}_\mathcal{I} (i)}_{\text{interference}} + \bm{n}_k(i),
\end{align}
where $\mathcal{S}_k^C$ is the complement of set $\mathcal{S}_k$ in $\mathcal{S}$. Let $\Pi_{\mathcal{S}_k}$ denote the collection of all non-empty subsets of $\mathcal{S}_k$, with each element of $\Pi_{\mathcal{S}_k}$ denoted by $\pi_{\mathcal{S}_k}^j, ~j\in[2^{\binom{K-1}{t}}-1]$. We denote $\mathcal{S}(i) \subset \mathcal{S}$ as the subset of messages transmitted in time slot $i$, i.e., $\mathcal{T}\in\mathcal{S}(i)$ if $\bm{w}_\mathcal{T}(i) \neq \bm{0}$.

Note that each user may decode more than one message in each transmission slot.
From the capacity region of the associated Gaussian multiple access channel, following conditions must be satisfied for successful decoding of all the intended messages at user $k$, $k \in [K]$, at time slot $i$:
\begin{align}\label{eq:capacity_constraint}
\sum\limits_{\mathcal{T} \in \pi_{\mathcal{S}_k}^j} R^\mathcal{T}(i) \leq \frac{n_i}{n}~\!\text{log}_2 \bigg(1+\sum\limits_{\mathcal{T} \in \pi_{\mathcal{S}_k}^j}\gamma_k^\mathcal{T}(i) \bigg),\;\forall \pi_{\mathcal{S}_k}^j \in \Pi_{\mathcal{S}_k},
\end{align}
where $R^\mathcal{T}(i)$ is the rate of the message $\bm{s}_\mathcal{T}(i)$, and $\gamma_k^\mathcal{T}(i)$ is the received signal-to-interference-plus-noise ratio (SINR) of message $s_\mathcal{T}(i)$ at user $k$ at time slot $i$, given by
\begin{align}
\gamma_k^\mathcal{T}(i) = \frac{\vert \bm{h}_k^H \bm{w}_\mathcal{T}(i) \vert^2}{\sum\nolimits_{\mathcal{I} \in \mathcal{S}_k^C} \vert \bm{h}_k^H \bm{w}_\mathcal{I}(i) \vert^2+\sigma_k^2},
\end{align}
for any $ \mathcal{T} \ni k $, or equivalently, any $\mathcal{T}\in \mathcal{S}_k$.
The rate of message $\mathcal{T}$ is the sum of the rate of submessages $s_\mathcal{T}(i)$, and must satisfy
% given by
\begin{align}
    % \tilde{R}^\mathcal{T}\triangleq
    \sum\nolimits_{i=1}^B R^\mathcal{T} (i) \geq \frac{R}{\binom{K}{t}},~\forall \mathcal{T}.
\end{align}

\subsection{Total Transmit Power Minimization}

To characterize the minimum power consumption in (\ref{AveragePowerMemoryTradeOff}) under the multi-antenna framework, a power minimization problem is formulated under the requirement of successfully delivering all the desired files to each user at the prescribed common rate.
For any given transmission scheme represented by the rates of the subfiles transmitted in each time slot $\{R^\mathcal{T}\!(i)~\!\!|\!\!~\forall~\!\!\mathcal{T}~\!\!\!\!\in\!\!\!\!~\mathcal{S}\}_{i=1}^B$ and the blocklength of each time slot $\{n_i\}_{i=1}^B$, the minimum average required power
can be obtained as
\begin{align}
P = \sum_{i=1}^B \frac{n_i}{n} P_i,
\end{align}
where
\begin{subequations}\label{eq:optimization}
\begin{flalign}
P_i \triangleq &\min\limits_{\{\bm{w}_\mathcal{T}(i)\}} \ \ \!\!\!\! \sum\limits_{\mathcal{T} \in \mathcal{S}} \ \Vert\bm{w}_{\mathcal{T}}(i)\Vert^2\\
&~~\text{s.t.} \ \ \!\!\!\! \sum\limits_{\mathcal{T} \in \pi_{\mathcal{S}_k}^j}\!\!\!\!R^\mathcal{T}(i) \!\leq\! \frac{n_i}{n}~\!\text{log}_2 \!\bigg(1\!+\!\!\!\!\sum\limits_{\mathcal{T} \in \pi_{\mathcal{S}_k}^j}\!\!\!\!\gamma_k^\mathcal{T}(i) \bigg),\;\!\!\forall \pi_{\mathcal{S}_k}^j \in \Pi_{\mathcal{S}_k},~\!\!\forall k.\label{eq:constraints}
\end{flalign}
\end{subequations}

The optimization problem in \eqref{eq:optimization} is nonconvex due to the constraints in (\ref{eq:constraints}); hence, it is computationally intractable.
In order to obtain an effective feasible solution of the problem (\ref{eq:optimization}), we leverage the SCA algorithm, which is known to converge to a stationary point of the original nonconvex problem \cite{Scutari}.
The details of the implementation of SCA are omitted, and will be elaborated in the longer version.

\section{Low-Complexity Delivery scheme}\label{ProposedSchemeSec}

We propose a content delivery scheme with the flexibility to adjust the number of coded messages intended for each user at each time slot.
Assuming a set $\mathcal{S}(i) = \{ \mathcal{T}~\!\!|\!\!~R^\mathcal{T}(i) \neq 0 \}$ of messages are transmitted in time slot $i$,
$c_k(i) \triangleq | \mathcal{S}(i)  \bigcap  \mathcal{S}_k| $ messages are transmitted to user $k$, then we have $2^{c_k(i)}-1$ constraints for user $k$ in time slot $i$ in problem (\ref{eq:optimization}).
Computational complexity increases drastically with the number of constraints, rendering the numerical optimization problem practically infeasible. In addition, a multi-user detection scheme, e.g., successive interference cancellation (SIC), needs to be employed at the users, whose complexity also increases with $c_k(i)$.

A low complexity scheme is proposed in \cite{DBLP:journals/corr/abs-1711-03364} by limiting the number of users to be served in each time slot, thereby indirectly reducing the number of coded messages to be decoded by each user. Instead of limiting the subsets of users to be served in each time slot, we propose to directly adjust the number of coded messages targeted to each user.
We will show that this results in a more efficient delivery scheme than the one in \cite{DBLP:journals/corr/abs-1711-03364}.

In the setting of our Example, if we transmit all the messages in one time slot, i.e., $B=1$, a total of $|\mathcal{S}|=\binom{K}{t+1}=10$ coded subfiles are transmitted simultaneously, with each user decoding $\binom{K-1}{t}=4$ messages. Accordingly, in the optimization problem in (\ref{eq:optimization}) we will have $K \times (2^{|\mathcal{S}_k|}-1) = 75$ constraints.
To alleviate the computational complexity, the low complexity scheme in \cite{DBLP:journals/corr/abs-1711-03364} splits each subfile into $3$ minifiles, and the coded messages are grouped to serve a subset of $K_s = 3$ users in each of the $B=\binom{K}{K_s}=10$ time slots. Within each time slot, each user needs to decode 2 messages to recover the corresponding minifiles.
Note that the power minimization problem for each time slot can be solved independently; therefore, we would need to solve 10 smaller optimization problems, each with $3\times 3=9$ constraints.

In contrast, we propose to serve as many users as needed at each time slot while keeping $c_k(i)$ under a given threshold $s$ for each user $k$. In our Example, we can satisfy all the user requests in only $2$ time slots, by setting nonzero rate targets for the messages in
\begin{align}
\mathcal{S}(1) &= \{{\{1,2\},\{2,3\},\{3,4\},\{4,5\},\{1,5\}}\}, \mbox{ and}\nonumber\\
\mathcal{S}(2) &= \{{\{1,3\},\{2,4\},\{3,5\},\{1,4\},\{2,5\}}\}\nonumber
\end{align}
in time slots 1 and 2, respectively. Note that each user $k \in [5]$ decodes only $c_k(i)=s= 2 $ messages in time slot $i=1,2$, the same as the delivery scheme in \cite{DBLP:journals/corr/abs-1711-03364}, requiring the same implementation complexity at each user; however, $5$ users are served in each time slot, which results in a significantly smaller number of time slots. Thus, we need to solve only two optimization problems at the BS, each with $5 \times 3=15$ constraints.

\begin{algorithm}[H]
\caption{Reduced-complexity delivery scheme}
\label{alg:loop}
\begin{algorithmic}[1]
\Require{$N,K,M,s,R$} 
\Ensure{$B$,~$\bigcup_{i=1}^B \{R^\mathcal{T}(i)\},\forall \mathcal{T}$}
% \Statex
\State Set $t=\frac{MK}{N}$, $i=1$, and $\mathcal{E}=\mathcal{S}$

    \While {$\mathcal{E}\neq \varnothing$}
    	\State Set $\bm{c}(i)\triangleq [c_1(i)\cdots c_{K}(i)]=\bm{0}$, $\mathcal{S}(i)=\varnothing$, $\mathcal{C}=\mathcal{E}$
    	\While {$c_k(i) \leq s,\forall k\in[K]$ and $\mathcal{C}\neq \varnothing$}
    		\State $\mathcal{K}\triangleq \{ k~\!|\!~\argmin\limits_k \bm{c}(i)\}$
    		\State Find $\hat{\mathcal{T}}=\argmax\limits_{\mathcal{T}\in \mathcal{C}} |\mathcal{K}\bigcap\mathcal{T}|$
    		\State $\mathcal{C} = \mathcal{C} \backslash \mathcal{\hat{T}}$
    		\If{$c_k(i)+1 \leq s, \forall k\in\mathcal{\hat{T}}$}
    		    \State $c_k(i) = c_k(i)+1, \forall k\in\hat{\mathcal{T}}$
    			\State $\mathcal{S}(i) = \mathcal{S}(i) \bigcup \mathcal{\hat{T}}$, $\mathcal{E} = \mathcal{E} \backslash \mathcal{\hat{T}}$
    		\Else \State \textbf{break}
    		\EndIf
    	\EndWhile
    	\State $i\leftarrow i+1$
    \EndWhile
\State Set $B = i-1$
\For {$i=1:B$}
    \State $n_i=\frac{|\mathcal{S}(i)|}{\binom{K+1}{t}}n$
    \State  \begin{subequations}
            $R^\mathcal{T}\!(i)=\!
            \begin{cases} \frac{R}{\binom{K}{t}},
             ~\forall~\!\mathcal{T}\in\mathcal{S}(i)\\
            0,~\text{otherwise}
            \end{cases}$
            \end{subequations}
\EndFor
\end{algorithmic}
\label{alg:proposed}
\end{algorithm}

In general, the number of constraints in the optimization problem in \eqref{eq:optimization} increases exponentially with $s$, which results in exponentially increasing number of constraints in the problem in each SCA iteration. Thus the computational complexity of the delivery scheme can be largely alleviated by choosing a small $s$ value, which also simplifies the multi-user detection algorithm.

In our proposed low-complexity scheme, we seek to
divide set $\mathcal{S}$ into disjoint subsets $\mathcal{S}(1),\cdots,\mathcal{S}(B)$, with $c_k(i)\leq s$ for $\forall k,i$, while keeping $B$ as small as possible.
To this end, we propose a greedy algorithm to construct the disjoint sets for any $s$ value, as shown in Alg.~\ref{alg:proposed}. Specifically, $\mathcal{S}(i)$'s are generated in a sequential manner:
to construct $\mathcal{S}(i)$, given the initialized $\bm{c} (i) \triangleq[c_i(i) \cdots c_K(i)] = \bm{0}$, $\mathcal{S}(i)=\varnothing$, and the set $\mathcal{E} = \mathcal{S}\backslash \bigcup_{j=1}^{i-1}\mathcal{S}(j)$ of remaining messages for assignment, user(s) that decode the least number of messages, i.e., user(s) in set $\mathcal{K} \triangleq \{ k~\!\!|\!\!~\argmin_k \bm{c}(i)\}$, are firstly identified, followed by checking whether there exists a message $\mathcal{\hat{T}}\in\mathcal{E}$ such that the condition $c_k(i)+1 \leq s$ for $\forall k \in \mathcal{\hat{T}}$ holds.
If such $\mathcal{\hat{T}}$ can be found, then $\mathcal{\hat{T}}$ is added to $\mathcal{S}(i)$ and removed from $\mathcal{E}$, and $\bm{c}(i)$ is updated by replacing $c_k(i) \leftarrow c_k(i)+1$, $\forall k \in \mathcal{\hat{T}}$. We initialize $\mathcal{C}=\mathcal{E}$ as the set of remaining messages in $\mathcal{E}$ for checking with the condition, and set $\mathcal{C}=\mathcal{C} \backslash \mathcal{T}$ each time after a message $\mathcal{T}\in\mathcal{E}$ is checked.
If a qualified $\mathcal{\hat{T}}$ is not found until $\mathcal{C}=\varnothing$, i.e., adding any of the messages in $\mathcal{E}$ to $\mathcal{S}(i)$ will violate the above condition, the process of constructing $\mathcal{S}(i)$ is completed, and we start constructing $\mathcal{S}(i+1)$ in the same manner.
The whole procedure is completed when $\mathcal{E} = \varnothing$, i.e., all the messages have been assigned to a subset. For efficient wireless resource management, we allocate $|\mathcal{S}(i)| n \slash \binom{K}{t+1}$ channel uses to transmit the messages in $\mathcal{S}(i)$, for $i \in [B]$.
Note that our proposed scheme covers the case of $B=1$, where all the messages are sent simultaneously, if $s \geq \binom{K-1}{t}$.

\section{Simulation Results}
In this section, numerical results are presented to demonstrate the effectiveness of the proposed coded delivery scheme with multi-antenna beamforming. We consider a single-cell with radius 500m, and users uniformly randomly distributed in the cell. Channel vectors $\bm{h}_{k}$ are written as $\bm{h}_k=(10^{{-\text{PL}}/10})\bm{\tilde{h}}_k$, $\forall k$, where $\bm{\tilde{h}}_k$ denotes an i.i.d. vector accounting for Rayleigh fading of unit power, and the path loss exponent is modeled as $\text{PL}=148.1+37.6\text{log}_{10}(v_k)$, with $v_k$ denoting the distance between the BS and the user (in kilometers). The noise variance is set to $\sigma^2_k=\sigma^2=-134$ dBW for all the users.
% We set the same rate target $R$ for each file for all the schemes.
All simulation results are averaged over 1000 independent trials computed with MOSEK in CVX \cite{cvx}.
We note that in simulations of our proposed low-complexity scheme, the blocklength of each time slot is set proportional to the number of messages transmitted in that time slot, and better performance in terms of average transmit power is observed compared to setting equal blocklengths for all the time slots.

\begin{figure}[!tp]
\centering
\includegraphics[width=2.65in]{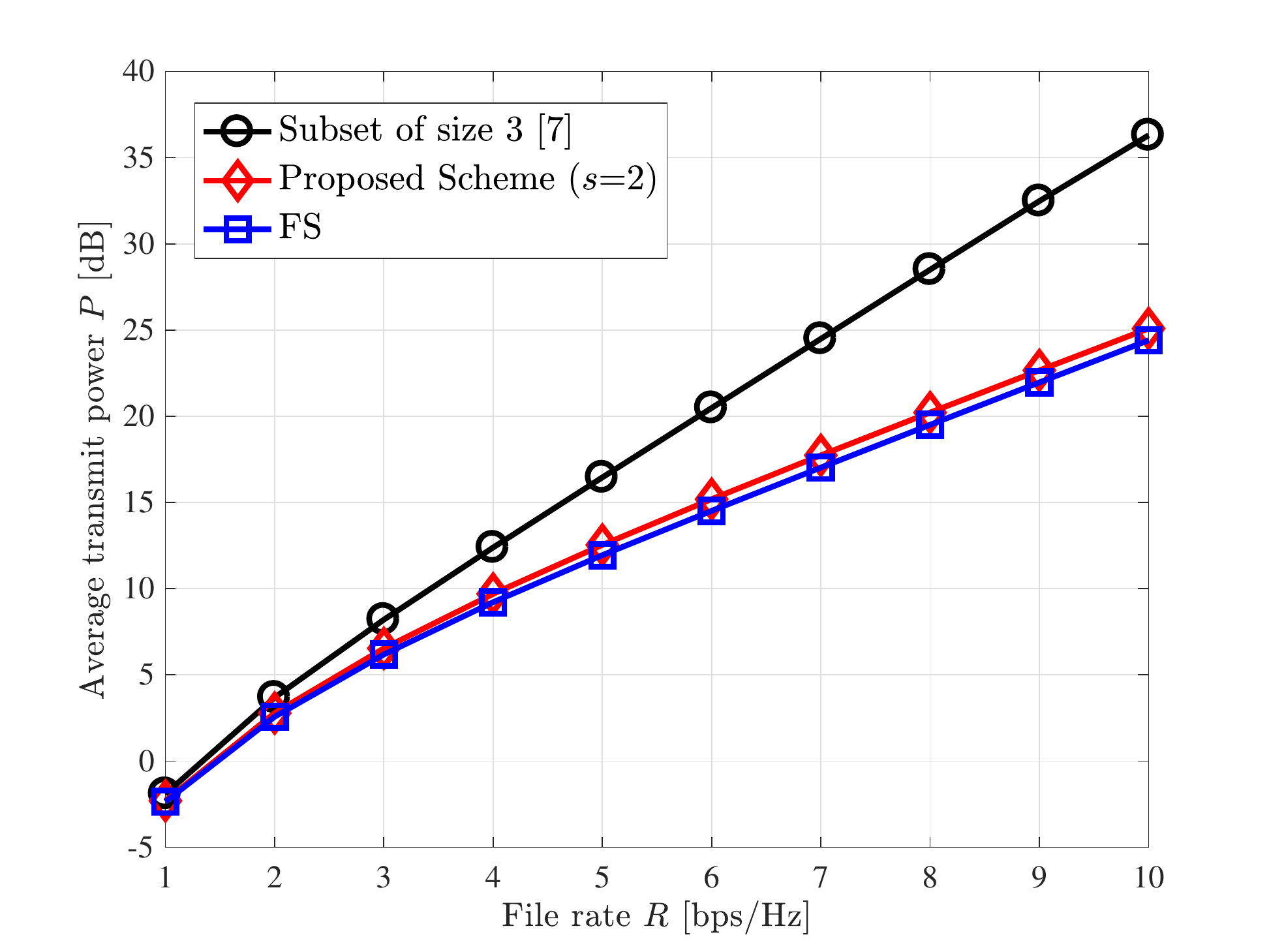}
\caption{Average transmit power $P$ as a function of the rate target $R$ for the network with $N=K=5$, $M=1$, and $N_t=6$.}
\label{PvR1}
\end{figure}

The scheme with $B=1$ time slot will be referred to as the full superposition (FS) scheme. FS has the best performance in terms of transmit power given enough number of spatial degrees of freedom, and serves as a baseline, but it also has the highest complexity.
To compare our results with those in \cite{DBLP:journals/corr/abs-1711-03364}, same number of coded messages are transmitted to each user in each time slot for both schemes. We note here that the scheme in \cite{DBLP:journals/corr/abs-1711-03364} can be improved by serving disjoint subsets of users simultaneously without increasing the complexity, but the improvement is only applicable when the size of user subset $K_s$ divides $K$. Therefore, the scheme in \cite{DBLP:journals/corr/abs-1711-03364} cannot handle certain settings such as the one in the running Example.

We first present the average transmit power as a function of the target rate $R$ in Fig.~\ref{PvR1} for the setting in our Example, assuming that the BS is equipped with $N_t=6$ antennas.
The scheme in \cite{DBLP:journals/corr/abs-1711-03364} that satisfies $s=2$ is adopted for fair comparison, where $K_s=3$ users are served in each time slot.
We observe that the proposed scheme provides significant savings in the transmit power compared to the one in \cite{DBLP:journals/corr/abs-1711-03364} at all rates.
The power savings increase with the target rate $R$ as a result of the increased superposition coding gain. Furthermore, the gap between the proposed scheme and FS is quite small, and remains almost constant with rate. At $R=10~ \text{bps/Hz}$, the power loss of the scheme in \cite{DBLP:journals/corr/abs-1711-03364} and ours compared to FS is about 12 dB and 1 dB, respectively.
Hence, the proposed scheme provides significant reduction in the computational complexity without sacrificing the performance much.

\begin{figure}[!tp]
\centering
\includegraphics[width=2.65in]{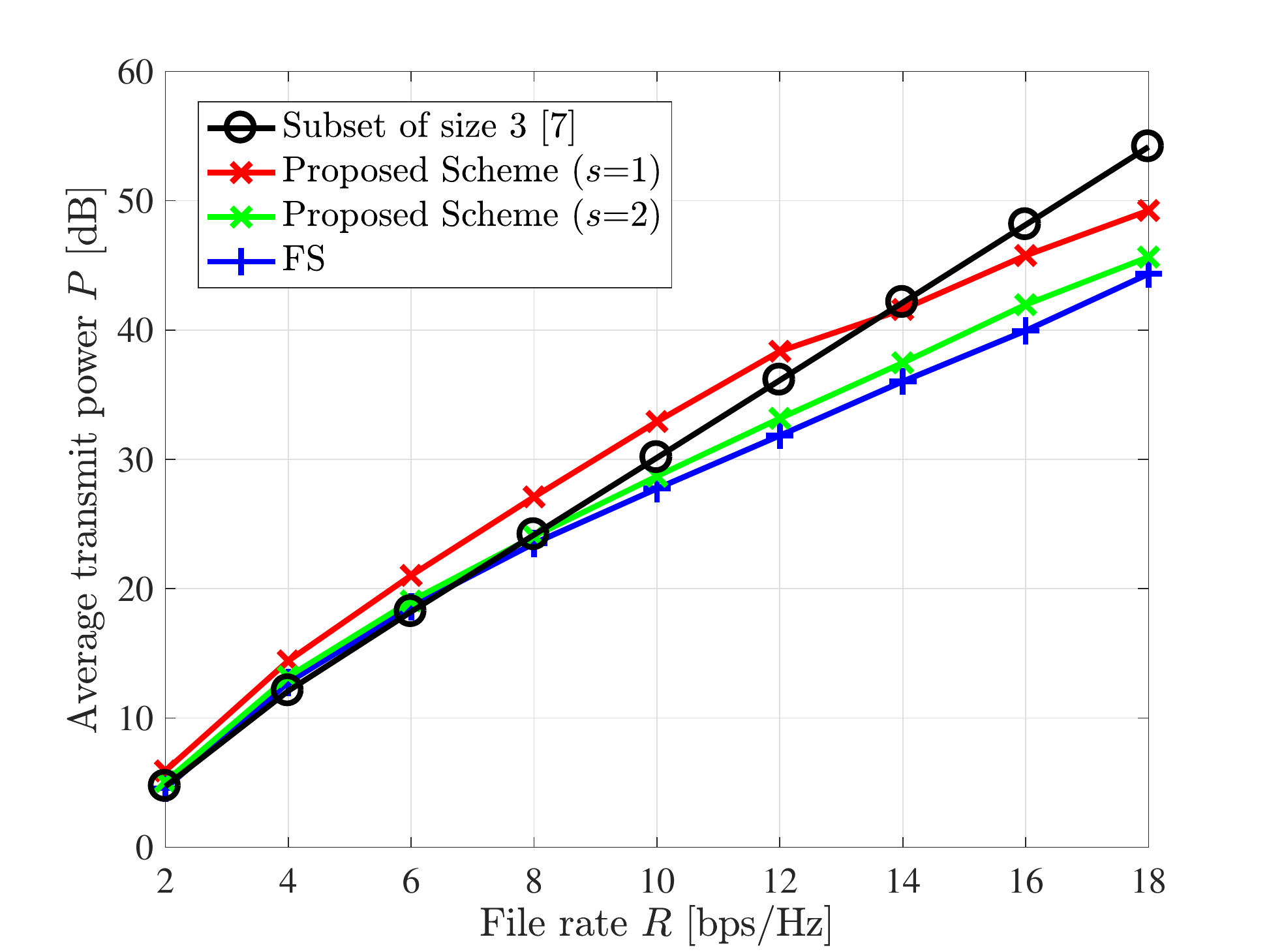}
\caption{Average transmit power $P$ as a function of the rate target $R$ for the network with $N=K=4$, $M=1$, and $N_t=3$.}
\label{PvR2}
\end{figure}

Fig.~\ref{PvR2} considers the average transmit power versus the rate target for $N=4$ files, $K=4$ users, $M=1$, and $N_t=3$ antennas. It is interesting to see that when the target rate is low, the scheme in \cite{DBLP:journals/corr/abs-1711-03364} slightly outperforms both the FS and the proposed schemes.
Due to insufficient spatial degrees of freedom, both the FS and the proposed schemes fail to manage the interference between data streams.
We conclude that this effect occurs only for low rates, as the benefit of superposition coding becomes more dominant at high rates.
We note that the performance of our scheme coincides with that of the scheme in \cite{DBLP:journals/corr/abs-1711-03364} for $s=1$, but this does not always happen. For instance, when $s=2$, the only option in \cite{DBLP:journals/corr/abs-1711-03364} to keep the same level of complexity is to serve 3 users in each time slot.

\begin{figure}[!tp]
\centering
\includegraphics[width=2.65in]{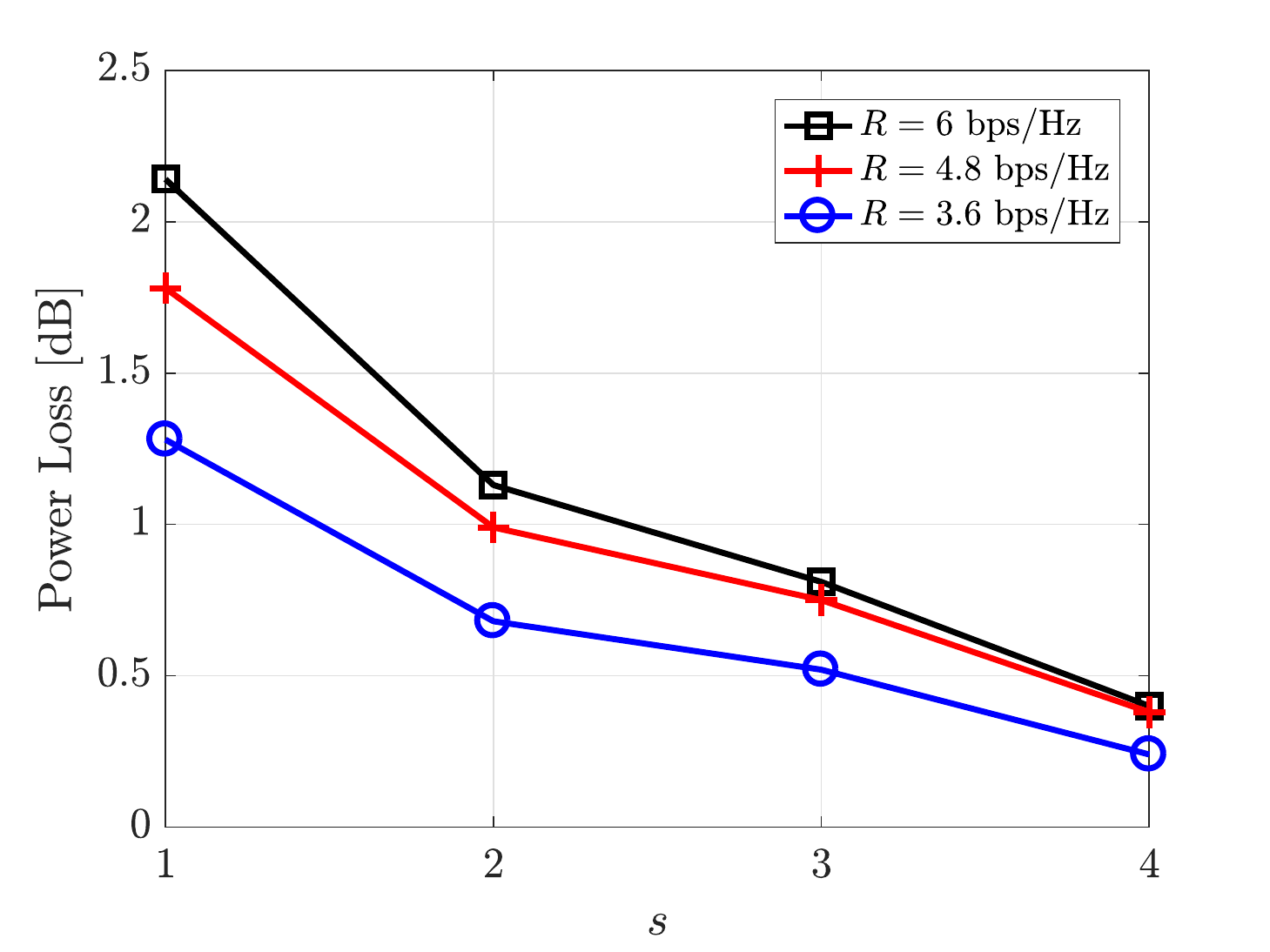}
\caption{Power loss with respect to FS as a function of $s$ with $N=K=6$, $M=1$, and $N_t=6$.}
\label{fig:power_loss}
\end{figure}

We finally plot in Fig.~\ref{fig:power_loss} the power loss of the proposed scheme compared to FS as a function of $s$. Assuming $N=6$, $K=6$, $M=1$, we let $s$ take values from $[5]$, where $s=5$ corresponds to the FS scheme in which all the $\binom{K}{t+1}=15$ coded messages are transmitted in one time slot.
At the other extreme, when $s=1$, the model boils down to the single-cell multigroup multicasting problem, which has the lowest computational and implementation complexity. In general, Fig.~\ref{fig:power_loss} can be considered as the trade-off curve between the performance and complexity for each target rate value, both of them increasing with $s$.

\section{Conclusions}

We have proposed a low-complexity coded content delivery scheme for a multi-antenna BS serving cache-enabled users.
By limiting the number of coded messages targeted at each user in each time slot, the proposed coded content delivery scheme provides the flexibility to adjust the computational complexity of the optimization problem and the receiver complexity. Compared with the FS scheme, in which all the coded messages are delivered simultaneously, our proposed delivery scheme achieves comparable performance with significantly lower complexity by delivering the coded messages in a time division fashion.
It has also been shown that the proposed delivery scheme outperforms the one proposed in \cite{DBLP:journals/corr/abs-1711-03364} for all values of SNR and rate with sufficient spatial degrees of freedom, while the improvement is limited to high data rate values when the BS does not have sufficiently many transmit antennas.
When considering practical implementations, one must choose a suitable value of $s$ that yields an acceptable performance while keeping the complexity feasible.

\bibliographystyle{IEEEtran}
\bibliography{IEEEabrv,bibtex.bib}

\end{document}